\documentclass[11pt]{article}
\usepackage{graphicx,tabularx,epsfig}
\usepackage{color}
\usepackage{enumitem}
\usepackage{caption}
\usepackage{subcaption}
\usepackage{amsmath}
\usepackage{amssymb}
\usepackage{amsthm}
\usepackage{physics}
\usepackage{dsfont, mathtools}
\usepackage{psvectorian}
\usepackage{blkarray}
\usepackage{bm}
\usepackage{url}
\usepackage{setspace}

\usepackage{xcolor}
\colorlet{linkequation}{blue}

\usepackage[colorlinks = true,
            linkcolor = blue,
            urlcolor  = blue,
            citecolor = blue,
            anchorcolor = blue]{hyperref}

\usepackage{floatrow}

\newlength{\abstractwidth}
\renewcommand{\thefootnote}{\fnsymbol{footnote}}
\renewcommand{\thanks}[1]{\footnote{#1}} 
\newcommand{\starttext}{
\setcounter{footnote}{0}
\renewcommand{\thefootnote}{\arabic{footnote}}}

\makeatletter
\g@addto@macro\normalsize{%
  \setlength\abovedisplayskip{15pt}
  \setlength\belowdisplayskip{15pt}
  \setlength\abovedisplayshortskip{15pt}
  \setlength\belowdisplayshortskip{15pt}
}
\makeatother

\usepackage{color}

\renewcommand{\title}[1]{\vbox{\center\LARGE{#1}}\vspace{5mm}}
\renewcommand{\author}[1]{\vbox{\center#1}\vspace{5mm}}
\newcommand{\address}[1]{\vbox{\center\em#1}}

\begin{document}

\singlespacing


\begin{center}

{\Large \bf {Size and momentum of an infalling particle\\ in the black hole interior}}
\end{center}

\bigskip \noindent

\bigskip

\begin{center}

\author{Felix M.\ Haehl and Ying Zhao}

\address{Institute for Advanced Study, Princeton, NJ 08540, USA}


 \vspace{0.5in}
    
    {\tt haehl@ias.edu, zhaoying@ias.edu}


\bigskip

\vspace{1cm}

\end{center}

\begin{abstract}


The future interior of black holes in AdS/CFT can be described in terms of a quantum circuit. We investigate boundary quantities detecting properties of this quantum circuit. We discuss relations between operator size, quantum complexity, and the momentum of an infalling particle in the black hole interior. We argue that the trajectory of the infalling particle in the interior close to the horizon is related to the growth of operator size. The notion of size here differs slightly from the size which has previously been related to momentum of exterior particles and provides an interesting generalization. The fact that both exterior and interior momentum are related to operator size growth is a manifestation of complementarity.

\medskip
\noindent
\end{abstract}

\newpage


\starttext \baselineskip=17.63pt \setcounter{footnote}{0}

{\hypersetup{hidelinks}
\tableofcontents
}

\section{Introduction}

In AdS/CFT, it was argued that the bulk geometry reflects the quantum circuit preparing the boundary state \cite{Swingle:2009bg,Hartman:2013qma,Susskind:2014moa}. In particular, the black hole interior has been associated with a unitary circuit preparing the state whose complexity corresponds to the volume / action of the interior spacetime \cite{Stanford:2014jda,Roberts:2014isa,Brown:2015bva}. It is then interesting to study properties of the interior in terms of the properties of the quantum circuit. 
 
\begin{figure}
 \begin{center}                      
      \includegraphics[width=.7\textwidth]{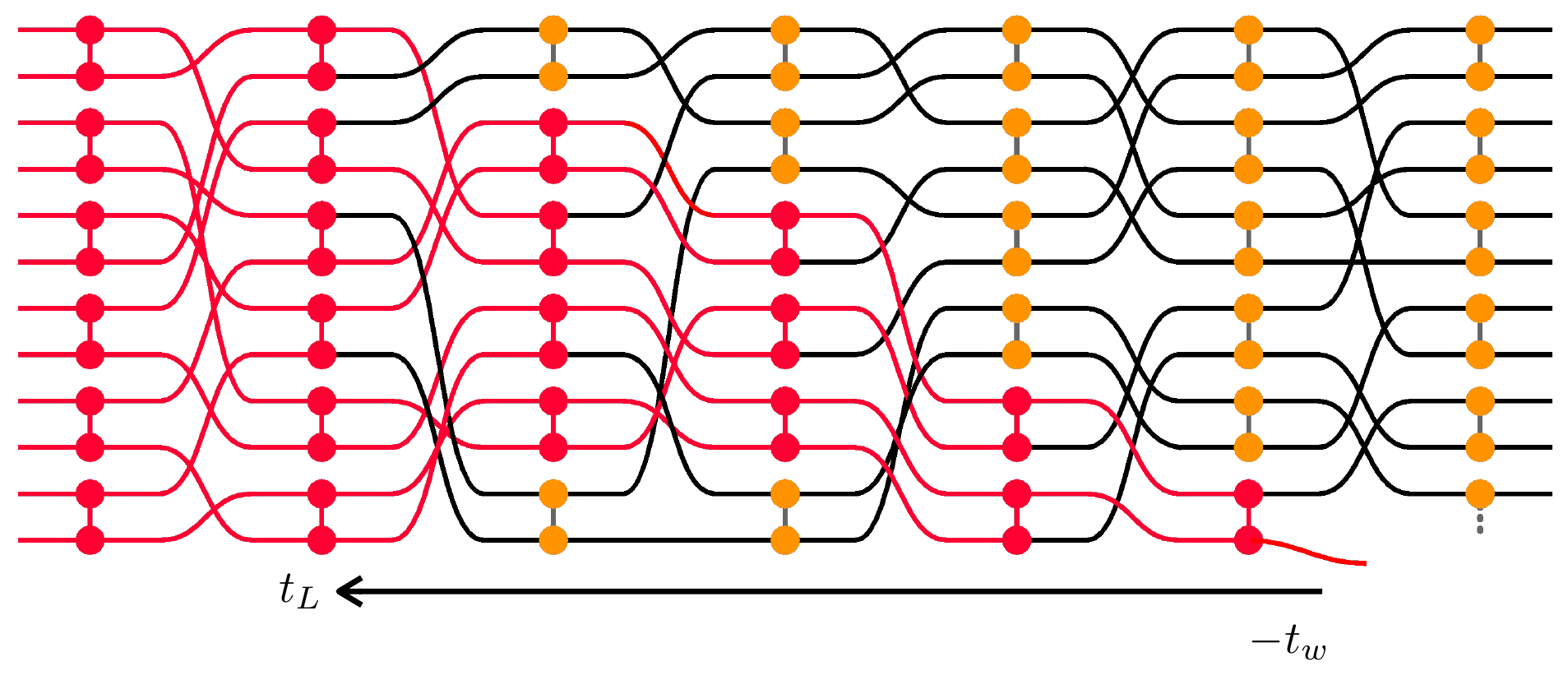}
      \caption{Illustration of the quantum circuit representing the future interior of the black hole dual to thermofield double perturbed from the left side.}
  \label{circuit_1_shock}
  \end{center}
\end{figure}

In this paper, we investigate boundary quantities detecting properties of the quantum circuit describing the interior. In particular, we imagine perturbing the thermofield double state from the left side. The perturbation acts like an ``infection'' due to adding one new qubit to the circuit, which propagates through the circuit by means of interactions in the quantum gates.\footnote{We perturb the black hole by applying a unitary operator. This will incerase the thermal entropy of the black hole. That's why we represent the perturbation by a new qubit in the circuit model.} This is shown in Figure \ref{circuit_1_shock}, which was argued to be the circuit representing the future interior of the bulk dual geometry \cite{Stanford:2014jda,Zhao:2017isy,Zhao:2020gxq}. We call gates to which the effect of the perturbation has spread as ``sick'' gates. What corresponds to the number of ``healthy'' gates per unit circuit time? It looks like the circuit in Figure \ref{circuit_1_shock} has to do with the exponential growth of a perturbation, so a natural guess is to consider an out-of-time-order four-point function like $\expval{\psi_1\psi_j(t)\psi_1\psi_j(t)}$. However, this four-point function is invariant under $t\rightarrow -t$. On the other hand, the perturbation in Figure \ref{circuit_1_shock} only grows in one direction. As the perturbation becomes small, why doesn't it grow toward the other direction?\footnote{We thank Xiaoliang Qi for asking this question.}

It was also argued that the absence of a firewall is tied to an increase of the quantum state's complexity with time \cite{Susskind:2015toa}. But the increase or decrease of complexity is not a state property since it depends on the Hamiltonian. On the other hand, we expect that the properties of the horizon are reflected by the quantum circuit in the interior. How are these related?

Finally, the increase of momentum of an infalling particle in the exterior was related to the growth of operator size \cite{Susskind:2018tei,Brown:2018kvn,Qi:2018bje}. If the relation between size and momentum is general, what corresponds to the momentum of the particle when it is in the interior? If we trace the particle backward from the singularity to the horizon, it again has exponentially increasing momentum. What can we say about this increase of momentum in terms of circuit properties and operator size?

We will show that the answers to these questions are related. We show that a particular four-point function counts the number of healthy gates in the interior circuit. Unlike $\expval{\psi_1\psi_j(t)\psi_1\psi_j(t)}$, this four-point function breaks time reflection symmetry $t\rightarrow -t$ and only grows toward one direction. We relate this four-point function to a notion of operator size which differs from the size that has been previously studied. We then relate this notion of operator size to the rate of change of complexity and comment on the connection to the firewall. We further relate complexity to momentum and show that this notion of size is related to the  momentum in the interior.

This paper is organized as follows. In section \ref{detection} we look for boundary quantities diagnosing properties of the quantum circuit stored in the interior. In section \ref{complexity} we discuss the relation between these circuit properties and the time dependence of complexity. In section \ref{size_momentum_in} we discuss the size/momentum relation in the black hole interior and comment on black hole complementarity. In section \ref{conclusion} we point out unanswered questions and future directions.

\section{Detecting properties of the interior circuit}
\label{detection}

In this section we discuss properties of the interior circuit and identify a simple quantity to probe them.

\subsection{Interior circuit corresponding to thermofield double}

We first look at the quantum circuit without perturbations. We represent the thermofield double by $S$ Bell pairs and model the dynamics by a Hayden-Preskill type circuit \cite{Hayden:2007cs}: at each time step the qubits are randomly grouped into $\frac{S}{2}$ pairs, and on each pair a randomly chosen $2$-qubit gate is applied.  As the left (right) boundary time increases, the circuit grows toward left (right). Correspondingly, the wormhole also grows.
\begin{figure}
 \begin{center}                      
      \includegraphics[width=.9\textwidth]{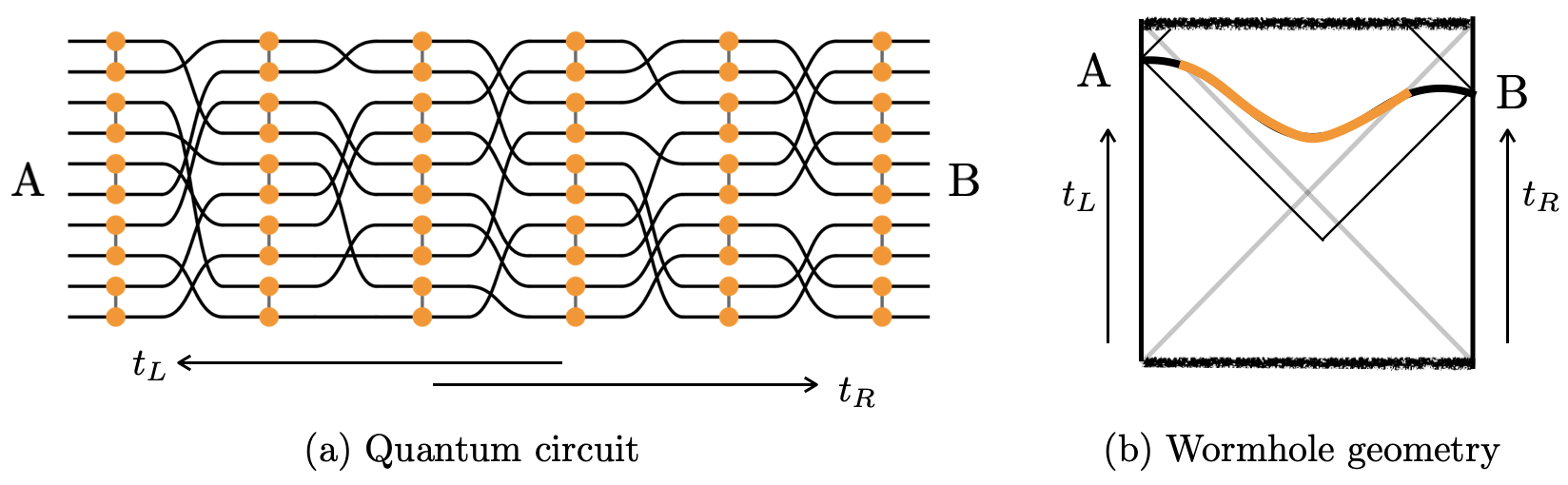}
      \caption{The size of the wormhole grows as the circuit grows.}
  \label{thermofield_double_time_evolved}
  \end{center}
\end{figure}
This is illustrated in Figure \ref{thermofield_double_time_evolved}. Note that the orange gates can be undone from both sides. We call them healthy gates. As we scan through the circuit, we can ask about the number of healthy gates per unit circuit time. In the case of the thermofield double it is a constant as the circuit is uniform in time.

\subsection{Interior circuit corresponding to perturbed thermofield double}

The epidemic model was introduced to describe the perturbation of a black hole \cite{Susskind:2014jwa,Roberts:2014isa}. For a black hole whose dynamics is modeled by a Hayden-Preskill type circuit, we can characterize the effect of some small perturbation as follows. Imagine the unperturbed system contains $S$ healthy qubits, and the perturbation is one extra qubit carrying some disease. The sick qubit enters the system at $\tau = 0$. Any qubits who interact directly or indirectly with sick qubits will get infected. We define the size of the epidemic $s_{ep}(\tau)$ to be the number of sick qubits at time $\tau$ where $\tau$ is discrete circuit time.\footnote{Later we will identify the circuit time and Schwarzshild time: $d\tau = \frac{2\pi}{\beta}dt$.} It satisfies
\begin{align}
&\frac{ds_{ep}(\tau)}{d\tau} = \frac{(S+1-s_{ep})s_{ep}}{S}\qquad\Rightarrow\qquad \frac{s_{ep}(\tau)}{S+1} = \frac{\frac{\delta S}{S}e^{\frac{S+1}{S}\tau}}{1+\frac{\delta S}{S}e^{\frac{S+1}{S}\tau}} \approx \frac{\frac{\delta S}{S}e^{\tau}}{1+\frac{\delta S}{S}e^{\tau}}\,.
\label{size_epidemic}
\end{align}
We used initial condition $s_{ep}(\tau = 0) = \delta S$ where $\delta S$ is the number of initially infected qubits, and $S+1$ is the size of the new circuit. $\delta S$ is proportional to the increase of thermal entropy from the perturbation. Figure \ref{circuit_1_shock} is the corresponding circuit picture. The orange dots represent the healthy gates (gates acting on healthy qubits) while the red dots represent sick gates (gates acting on sick qubits).


It was argued that a quantum circuit like the one in Figure \ref{circuit_1_shock} is stored in the future interior of the black hole corresponding to the perturbed thermofield double \cite{Zhao:2017isy,Zhao:2020gxq}. We briefly review the argument here. We imagine Alice and Bob share a thermofield double state. Alice holds the left side while Bob holds the right side. Alice applies a perturbation from the left at time $t_w$. 

\begin{figure}[H] 
 \begin{center}                      
      \includegraphics[width=\textwidth]{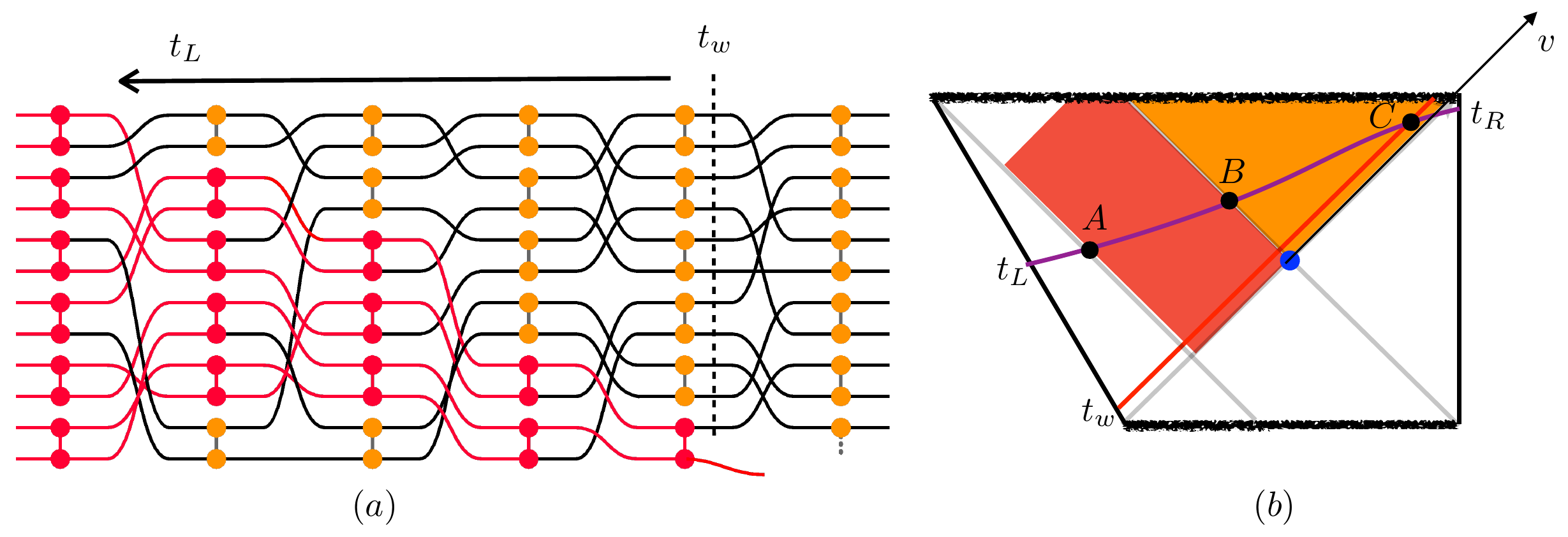}
      \caption{Illustration of the evolution of healthy and infected gates in (a) compared to the size of the associated bulk regions on time slices in (b).}
  \label{geodesic}
  \end{center}
\end{figure}

First, we look at the geometry from the point of view of Alice. We fix the right time at some large value and vary the left time (Figure \ref{geodesic}). As Alice varies the left time, the growth of the orange region on the time slices corresponds to the growth of the number of orange gates (healthy gates), while the growth of the red region corresponds to the increase in the red gates (sick gates). One can show this by comparing the geodesic contained in the red (orange) region and the number of sick (healthy) gates in the epidemic model \cite{Zhao:2017isy}.

\begin{figure}[H]
 \begin{center}                      
      \includegraphics[width=\textwidth]{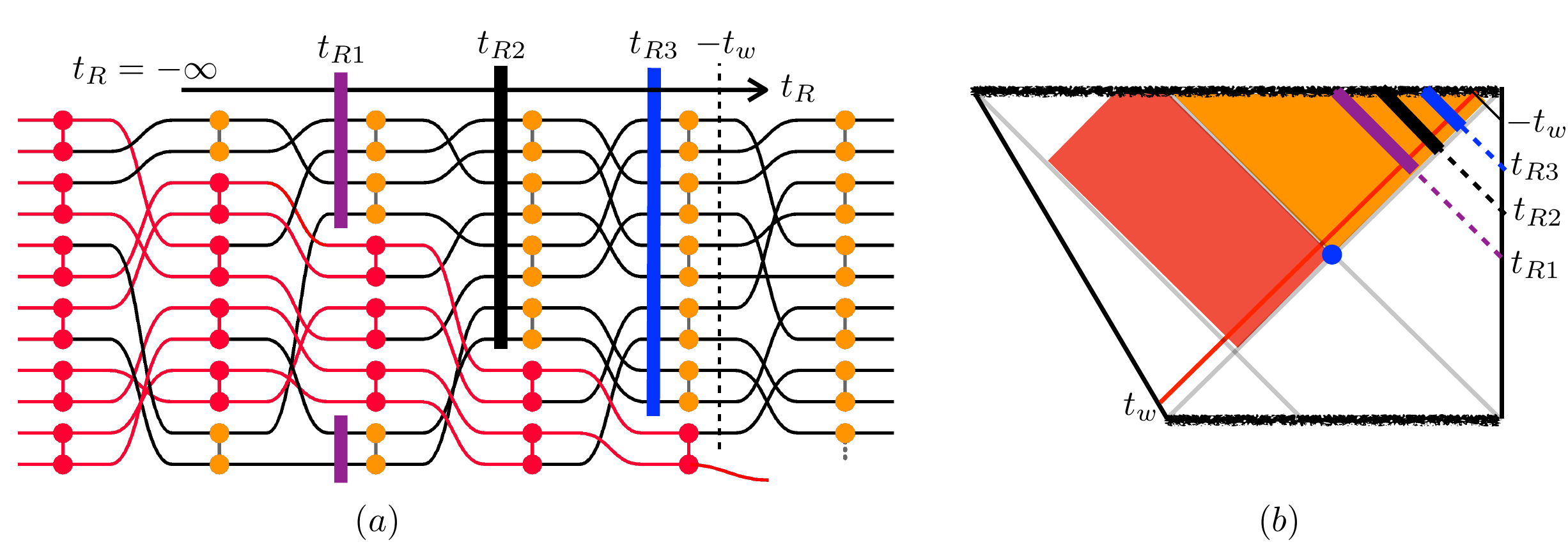}
      \caption{When Bob decreases the right time, he is scanning through the quantum circuit from the right to the left. }
  \label{Penrose_1_shock}
  \end{center}
\end{figure}

Next, we look at the geometry as well as the circuit from the point of view of Bob. We fix the left time at some large value and vary the right time (Figure \ref{Penrose_1_shock}). From the point of view of Bob, when he decreases the right time, he can undo the orange gates (healthy gates) stored in the circuit but cannot undo the red gates (sick gates). Instead, he will create a fold on top of the red gates.\footnote{We thank Juan Maldacena for pointing this out.} When he decreases the right time, his backward time evolution will cancel the healthy gates in the quantum circuit from the right to the left (Figure \ref{Penrose_1_shock}). This gives us some hint that the part of the trajectory close to the horizon has to do with the growth of the operator in the quantum circuit \cite{Zhao:2020gxq}.

\subsection{Counting the number of healthy gates}

\begin{figure}[H]
 \begin{center}                      
      \includegraphics[width=0.7\textwidth]{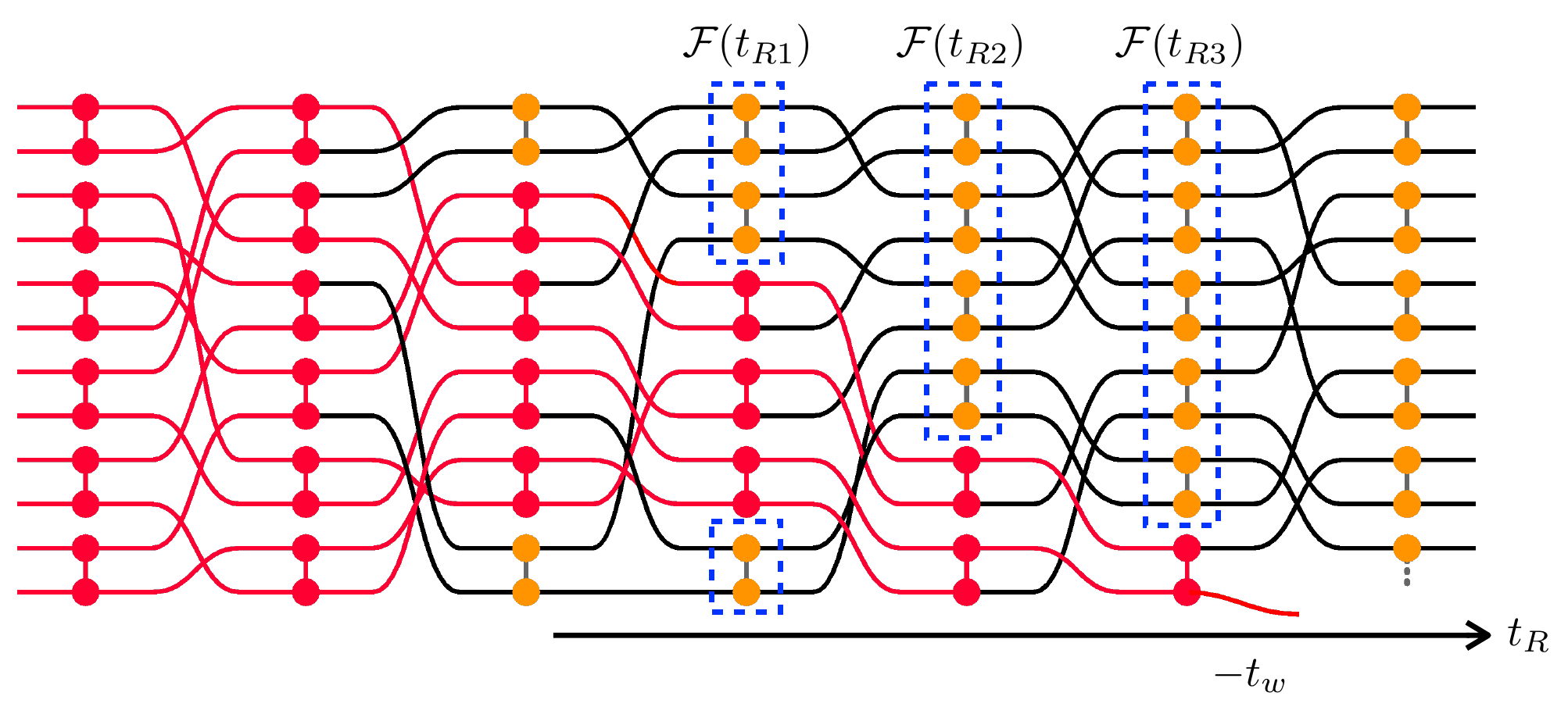}
      \caption{The number of healthy gates inside the blue dashed squares at time $t_{Ri}$ is given by ${\cal F}$.}
  \label{def_F}
  \end{center}
\end{figure}

If we say that a circuit like Figure \ref{circuit_1_shock} is stored in the future interior, what boundary quantities can we use to detect the properties of the circuit?  In particular, what boundary quantity gives the number of healthy gates in the circuit as a function of right time? In Figure \ref{def_F}, $\mathcal{F}(t_{Ri})$ denotes the fraction of healthy gates at time step $t_{Ri}$. 

To be concrete from now on we will work with the low energy limit of the SYK model described by JT gravity \cite{Kitaev:2015talk,Maldacena:2016hyu,Maldacena:2016upp}. We will use the convention that $\{\psi_i, \psi_j\} = 2\delta_{ij}$ so $\psi_i$ is unitary. 

The quantum circuit in Figure \ref{circuit_1_shock} clearly has to do with the growth of an operator, so the first guess would be to consider an out-of-time-order correlator like $\sum_{j}\expval{\psi_1\psi_j(t)\psi_1\psi_j(t)}$, where $\psi_1$ represents the perturbation and $\psi_j(t)$ detects its size at time $t$. However, this cannot quantify the kind of perturbation shown in Figure \ref{circuit_1_shock} because the quantity is invariant under $t\rightarrow -t$, while the perturbation in the circuit only grows towards the left. A related fact is that if Bob jumps in at right time $t_R>-t_w$, he will not meet the perturbation \cite{Zhao:2020gxq}.\footnote{This is not true for charged black holes.} If we consider the number of healthy gates as a function of right time $t_R$, it should first increase and then saturate.

So it is clear that we need some observable characterizing the growth of a perturbation, but the growth should be only in one direction. We consider the following quantity:
\begin{align}
\label{healthy_1}
	\mathcal{F}(a, t_R) = -\frac{\sum_{j = 1}^N\tr(\psi_1 \psi_j(-a)\psi_1\rho^{\frac{1}{2}}\psi_j(t_R)\rho^{\frac{1}{2}} )}{\sum_{j = 1}^N\tr(\psi_1\psi_1\rho)\tr(\psi_j(-a)\rho^{\frac{1}{2}}\psi_j(t_R)\rho^{\frac{1}{2}})}
\end{align}
Let's explain this quantity. We start from the thermofield double state $|\text{TFD}\rangle$ dual to eternal black hole geometry \cite{Maldacena:2001kr}, and perturb it from the left by applying $\psi_1$ at time $t_w = 0$. Then we look at the correlation between left time $a$ and right time $t_R$. In other words, we write \eqref{healthy_1} in the form\footnote{We assume the left time goes up in the following expression, c.f., Figure \ref{thermofield_double_time_evolved}.}
\begin{align}
\label{healthy_2}
	\mathcal{F}(a, t_R) = \frac{\sum_{j = 1}^N\bra{\text{TFD}}\psi_1^L\psi_j^L(a)\psi_j^R(t_R)\psi_1^L\ket{\text{TFD}}}{\sum_{j = 1}^N\bra{\text{TFD}} \psi_1^L \psi_1^L \ket{\text{TFD}} \bra{\text{TFD}}\psi_j^L(a)\psi_j^R(t_R)\ket{\text{TFD}}}
\end{align}
  \begin{figure} 
 \begin{center}                      
      \includegraphics[width=1.6in]{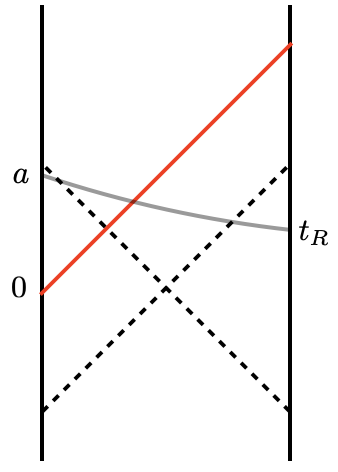}
      \caption{Illustration of the geodesics anchored at a late time $t_L = a$ on the left, thus serving as a probe of the interior.}
  \label{JT_correlation}
  \end{center}
\end{figure}
This is illustrated in Figure \ref{JT_correlation}.
Notice that this quantity is well suited to detect properties of the interior in the sense that the geodesic connecting left time $a$ and right time $t_R$ goes through the interior region for $a$ large and positive.

\begin{figure} 
 \begin{center}                      
      \includegraphics[width=4in]{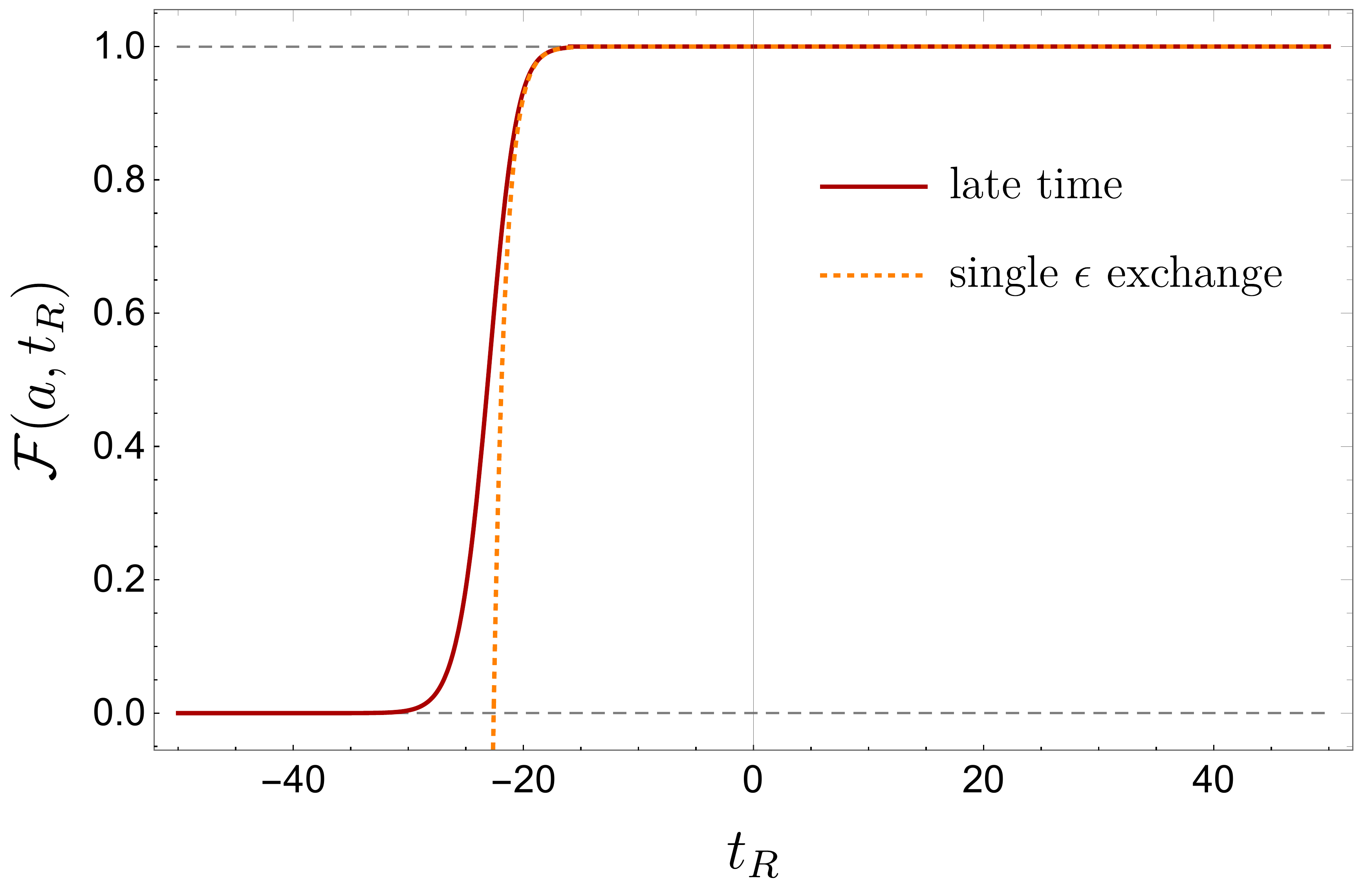}
      \caption{Plot of the four-point function ${\cal F}(a,t_R)$ defined in \eqref{healthy_1} and its leading approximation corresponding to the exchange of a single Schwarzian mode (see \eqref{eq:Fapprox}). We set $a = 100$, $\beta = 2\pi$, $(S-S_0) \sin \delta =10^{10}$, and $\Delta = \frac{1}{2}$. (Setting $\Delta = \frac{1}{2}$ gives the best match with the epidemic model and later bulk calculations.)}
  \label{FourPtPlot1}
  \end{center}
\end{figure}

Now we fix some large positive $a$, and consider $\mathcal{F}$ as a function of $t_R$.\footnote{ We take $t_*< a\ll \frac{\beta}{2\pi}(S-S_0)$, where $t_* = \frac{\beta}{2\pi} \log(S-S_0)$ is the scrambling time.} In the low energy limit of the SYK model we one can compute the Schwarzian contribution to the four-point function by resumming contributions for $t_R \sim \log (S-S_0)$ (see Appendix \ref{app:SchwCalc}). For generic operators with dimension $\Delta$, the result is:
\begin{align}
	{\cal F}(a,t_R) = z^{-2\Delta} \, U(2\Delta,1,\frac{1}{z}) \,,\qquad z = \frac{\beta}{8\pi C} \frac{1}{\sin \delta} \, \frac{e^{\frac{2\pi}{\beta}a}}{1+ e^{\frac{2\pi}{\beta}(a+t_R)}}
	\;\stackrel{a \gg -t_R}{\approx}\; \frac{\pi}{2(S-S_0)} \frac{1}{\sin \delta}\,e^{-\frac{2\pi}{\beta}t_R}
\end{align}
where we used $4\pi^2 CT = S-S_0$. We also have $\delta \sim \frac{\pi}{\beta\mathcal{J}}$.\footnote{We need to smear the operator insertions slightly in order to go from the UV to the conformal regime. The amount of smearing is $\sim\frac{1}{\mathcal{J}}$.}
Note that even if we consider finite large $a$, $z$ only grows for negative $t_R$ and does not grow for positive $t_R$. This will make sure that $\mathcal{F}$ only decays for negative $t_R$. We show a plot of ${\cal F}(a,t_R)$ in Figure \ref{FourPtPlot1}.

Using the limiting from of the hypergeometric function,
\begin{align}
	z^{-2\Delta}U(2\Delta, 1, \frac{1}{z}) \approx \begin{cases}
		1-4\Delta^2 z & z\ll 1\\
		\frac{1}{\Gamma(2\Delta)}z^{-2\Delta}\log z & z\gg 1
	\end{cases}
\end{align}
we have
\begin{align}
\label{percentage_sick}
	1-\mathcal{F}(a, t_R) \approx \begin{cases}
		\frac{2\pi\Delta^2}{(S-S_0)\delta}e^{-\frac{2\pi}{\beta}t_R} & -t_R<t_*\\
		1-\frac{1}{\Gamma(2\Delta)} e^{-2\Delta\frac{2\pi}{\beta}(-t_R-t_*)}(-t_R-t_*) & -t_R>t_*
	\end{cases}
\end{align}

We see that $1-\mathcal{F}$ behaves like the percentage of infected qubits in the circuit model. If we compare \eqref{percentage_sick} with \eqref{size_epidemic}, we see that $1-\mathcal{F}$ exhibits the exponential growth and saturation expected from the epidemic model once we identify $-\frac{2\pi}{\beta}t_R$ with circuit time $\tau$. 
Our proposal is that $\mathcal{F}(a, t_R)$ gives the percentage of healthy gates stored in the future circuit as a function of the right time $t_R$. The value of the parameter $a$ plays little role once we take it to be very large.

Further, note that when $t_R<-t_*$, the number of healthy gates is almost zero. When $t_R>-t_*$, the number of healthy gates is almost maximal. What this says is that the perturbation peels off from the horizon at right time $t_R = -t_*$.

\subsection{Growth of an operator in the interior circuit}

Next, we will provide an interpretation of ${\cal F}(a,t_R)$ in terms of the growth of the perturbing operator. We will first review some salient features of operator growth, mostly following \cite{Qi:2018bje}, and then discuss the relation to the epidemic model.

\subsubsection{Operator growth in the SYK model}

One can define the size of an operator ${\cal O}$ (made out of SYK fermions) by expanding it  in a basis of strings of elementary fermions $\psi_{j_1} \cdots \psi_{j_k}$ and identifying the number of fermions that occur in such strings. More precisely, we define a (positive and Hermitian) size operator $\hat{n}_\infty$ whose expectation value counts the average number of distinct flavors $j$ occurring in ${\cal O}$:
\begin{equation}
     \hat{n}_\infty = \frac{1}{2}\sum_j (1 + i \psi_j^L \psi_j^R): \qquad n_\infty[{\cal O}] = \langle {\cal O} | \hat{n}_\infty | {\cal O} \rangle =  \frac{1}{4}\sum_j \text{tr} \left( \{ {\cal O}, \psi_j \}^\dagger \{ {\cal O},\psi_j \} \right) \,.
\end{equation}
Here we have in mind a doubled Hilbert space such that $|{\cal O}\rangle \equiv  ({\cal O}_L \otimes \mathbf{1}_R) |0 \rangle$ is a state in ${\cal H}_L \otimes {\cal H}_R$, where $\ket{0}$ is the maximally entangled state.\footnote{ We use a convention where $\{\psi_i, \psi_j \} = 2 \delta_{ij}$. The maximally entangled state $|0\rangle$ satisfies $(\psi_j^L + i \psi_j^R) |0\rangle = 0$ for all $j=1,\ldots,N$. Note that $|0\rangle$ is also the thermofield double state at infinite temperature, hence the notation $\hat{n}_\infty$.}

The size of a single fermion $\psi_i$ is $n_{\infty}[\psi_i] = 1$. Similarly, the identity operator has size $n_{\infty} [\mathbf{1}] = \langle 0 | \hat{n}_\infty | 0\rangle = 0$. The peak of the size distribution describing a fully scrambled operator is $n_{max} = \frac{N}{2}$. This is the maximum size an operator can attain through scrambling. 

Consider now a thermal state at inverse temperature $\beta$, which we denote as $\rho$, and focus on the operator ${\cal O} = \psi_1(t)$.
In order to define its size, we use the purification of the state $\rho^{\frac{1}{2}}$, i.e., the thermofield double state, which we denote as $|\text{TFD}\rangle \in {\cal H}_L \otimes {\cal H}_R$. It was shown in \cite{Qi:2018bje} that the appropriate notion of size in this background is
\begin{equation}
\label{size_exterior}
   \frac{ n_{\beta}^\text{ex}[\psi_1(t)]}{n_{max}} \equiv \frac{n_\infty[\psi_1(t) \rho^{\frac{1}{2}}] - n_\infty[\rho^{\frac{1}{2}}]}{n_{max} - n_{\infty}[\rho^{\frac{1}{2}}]} \,,
\end{equation}
where $n_{max} = \frac{N}{2}$.
This quantity represents the average number of fermions in the operator $\psi_1(t)$ in a thermal background of $\rho^{\frac{1}{2}}$. We call it the ``exterior'' size because it is closely related to the momentum of the infalling particle in the exterior of the black hole, as we will discuss below.\footnote{ It is also well known that this quantity takes the form of a standard out-of-time-order four-point function:
\begin{equation}
    1 - \frac{n_\beta^\text{ex}[\psi_1(t)]}{n_{max}} = - \frac{\sum_j \text{tr} \left( \psi_1(t) \psi_j \psi_1(t) \, \rho^{\frac{1}{2}} \psi_j \rho^{\frac{1}{2}}\right)}{\sum_j \text{tr} \left(  \psi_j \, \rho^{\frac{1}{2}} \psi_j \rho^{\frac{1}{2}}\right)} \,.
\end{equation}
}


\subsubsection{Operator growth in the interior circuit}

Let us now return to the four-point function \eqref{healthy_1} and explain how it relates to operator growth in the thermal state. We claim that from the SYK point of view, $1-\mathcal{F}$ also characterizes the growth of the perturbation $\psi_1$ in the space of operators.


Indeed, from the definition of $\mathcal{F}$ in \eqref{healthy_1} one can show that
\begin{align}
\label{size_interior}
	 1-\mathcal{F}(a, -t)  =  \frac{n_{\infty}[e^{iH(a-t)}\psi_1(t)\rho^{\frac{1}{2}}]-n_{\infty}[e^{iH(a-t)}\rho^{\frac{1}{2}}]}{n_{max}-n_{\infty}[e^{iH(a-t)}\rho^{\frac{1}{2}}]}\equiv \frac{n_\beta^\text{in}[\psi_1(t)]}{n_{max}}\,.
\end{align}
The superscript ``in" means interior as we will later compare this with interior momentum.
We see that \eqref{size_interior} has the same form as the definition of size in \eqref{size_exterior}. One can consider it as the size of the operator $\psi_1(t)$ in the background $e^{iH(a-t)}\rho^{\frac{1}{2}}$.\footnote{As we mentioned before, the dependence of $\mathcal{F}$ on $a$ is negligible when $a$ is large. We can therefore approximate the background $e^{iH(a-t)}$ by a time independent expression $e^{iHa}$.\label{alternative_def}} A large positive $a$ breaks time-reflection symmetry and guarantees that this quantity only grows in one direction. 

We can gain some intuition about why size defined in \eqref{size_interior} only grows in one direction. In the numerator, we are comparing the size of $e^{iHa}\psi_1e^{-iHt}\rho^{\frac{1}{2}}$ and $e^{iHa}e^{-iHt}\rho^{\frac{1}{2}}$, with and without $\psi_1$ insertion.
Let us first note that the size of $e^{iHa}\psi_1e^{-iHt}\rho^{\frac{1}{2}}$ is essentially a four-point function of fermions. For $t>0$ this is an out-of-time-ordered correlation function, but it is time-ordered for $t<0$:
\begin{equation*}
    \includegraphics[width=.85\textwidth]{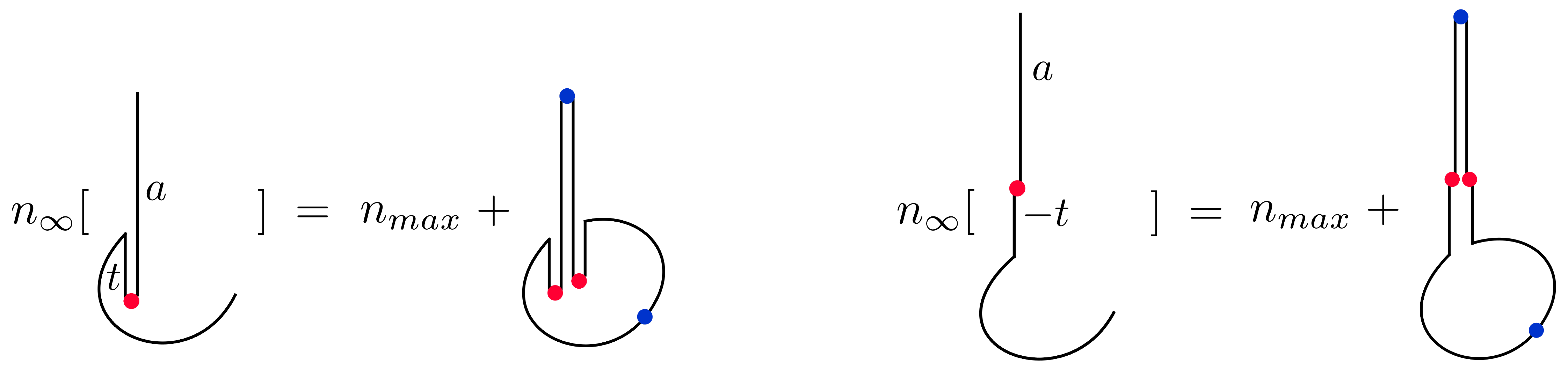}
\end{equation*}
where red dots indicate the insertions of $\psi_1(t)$, and blue dots denote the fermions of the size operator whose expectation value we compute.
This explains why the ``interior'' size grows exponentially for $t>0$, but does not grow for $t<0$. 

In other words, when $t>0$, the numerator of \eqref{size_interior} is roughly related to the size of a precursor operator, whose size grows with $t$:
\begin{equation*}
    \includegraphics[width=.95\textwidth]{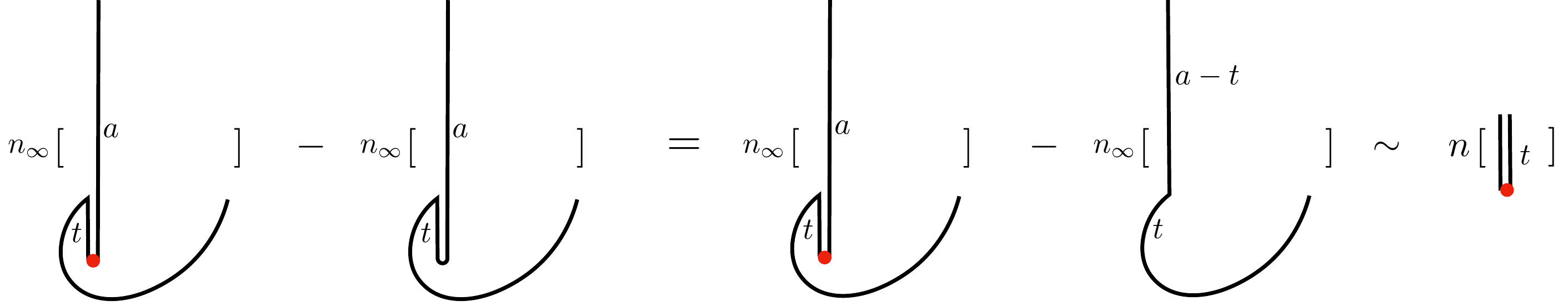}
\end{equation*}
On the other hand, when $t<0$, we simply get:
\begin{equation*}
    \includegraphics[width=2.5in]{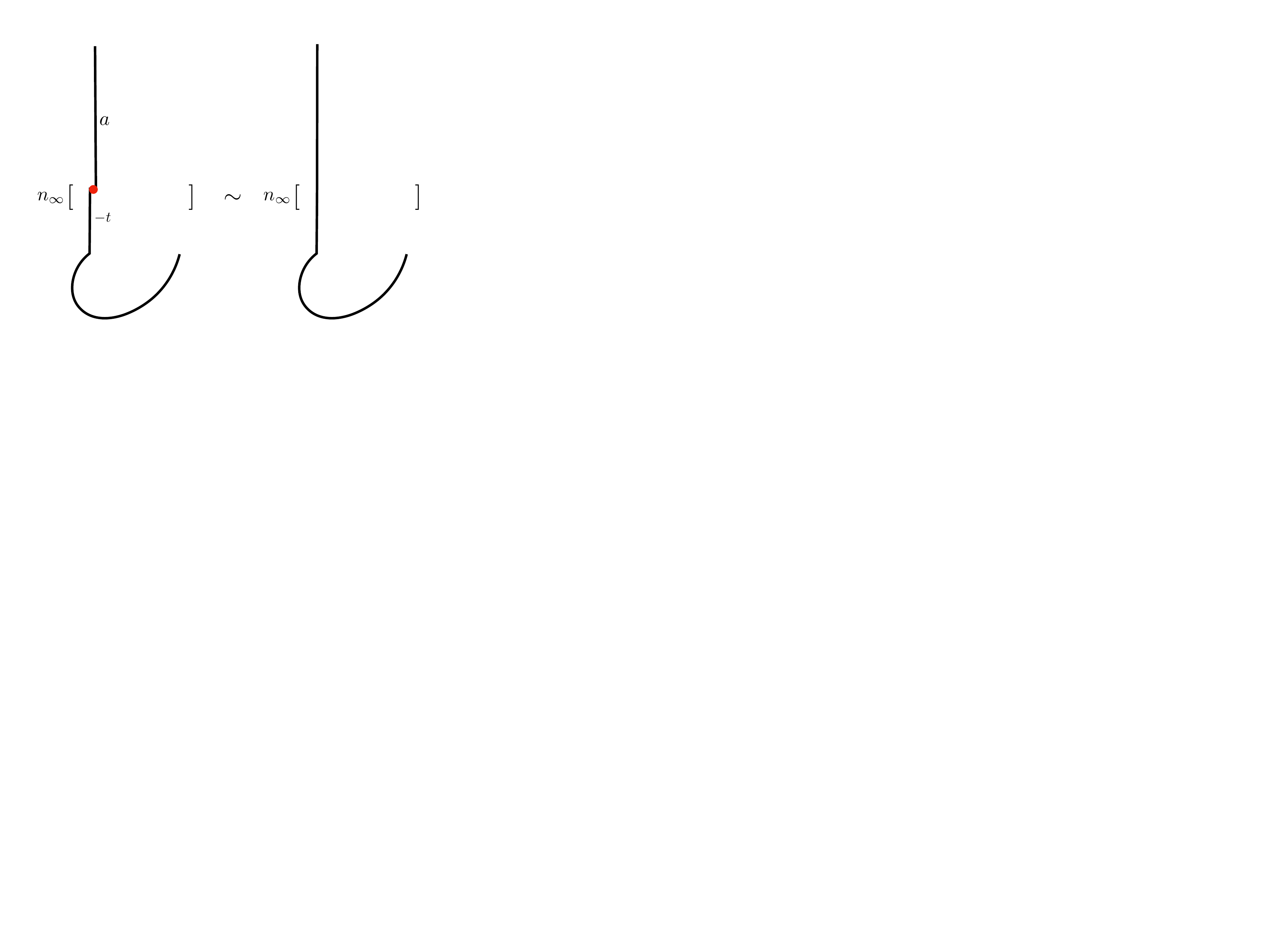}
\end{equation*}

If $\mathcal{F}$ gives the percentage of the number of healthy gates, it is also intuitive that $1-\mathcal{F}$ gives percentage of the number of sick gates. So $n_{max}(1-\mathcal{F})$ corresponds to the size of the epidemic in the circuit picture.

\section{Time dependence of complexity and properties of the interior circuit}
\label{complexity}

In \cite{Susskind:2015toa} Susskind argued that one can use time dependence of complexity to diagnose the smoothness of the horizon. On the other hand, we expect that properties of the horizon can be reflected as properties of the quantum circuit stored in the interior. So these two must be related. 

In fact, the setup used in the definition of the correlator $\mathcal{F}$ in \eqref{healthy_2} is the same as the setup used in \cite{Susskind:2015toa}. In both cases, one takes the left time to positive infinity and varies the right time. 

\begin{figure}[H] 
 \begin{center}                      
      \includegraphics[width=5.3in]{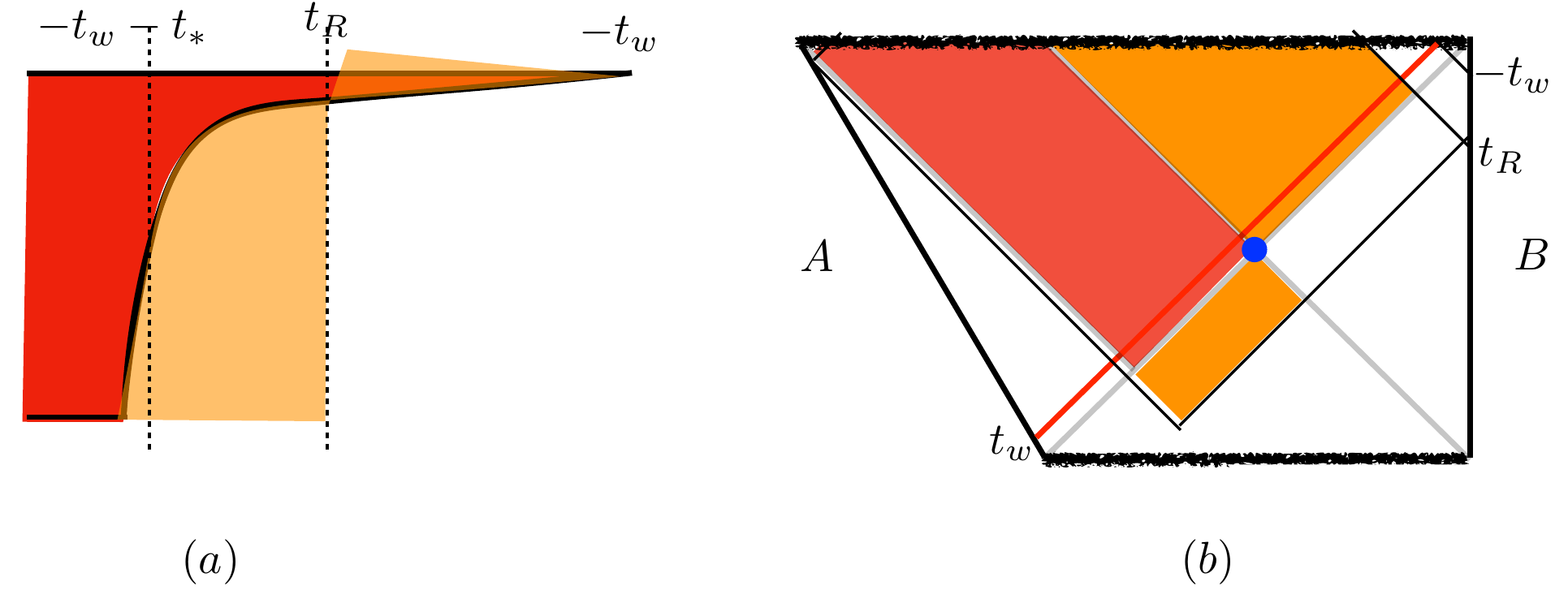}
      \caption{(a) Quantum circuit at $-t_w-t_*<t_R<-t_w$. As Bob decreases the right time, he will create a fold on top of the sick gates. (b) WDW patch at $-t_w-t_*<t_R<-t_w$.}
  \label{circuit_fold_1}
  \end{center}
\end{figure}

Say, in the circuit in Figure \ref{Penrose_1_shock}, Bob decreases the right time from $t_{R3}$ to $t_{R2}$. What is the corresponding change in complexity? Bob will undo the healthy gates while creating a fold on top of the sick gates (Figure \ref{circuit_fold_1}).\footnote{The fold in the circuit is the result of Bob implementing backwards time evolution: if after one time step, $S$ gates are applied, Bob can undo $S-s_{ep}$ of them. The remaining $s_{ep}$ have been affected by the perturbation and cannot be undone.}

The change of complexity as a function of $t_R$ is
\begin{align}
\label{complexity_perturbation}
	&\frac{d\mathcal{C}}{d(-t_R)} \propto T\qty[n_{max}(1-\mathcal{F})-n_{max}\mathcal{F}]
	=n_{max}T(1-2\mathcal{F}) \,,
\end{align}
where $n_{max}(1-\mathcal{F})$ is the number of gates added in the fold per unit time step, while $n_{max}\mathcal{F}$ is the number of canceled healthy gates per unit time step.
We see that $\frac{d\mathcal{C}}{dt_R}$ becomes negative when the percentage of healthy gates becomes small, or the percentage of sick gates gets large.

The interior spacetime stores a quantum circuit. From the point of view of the black hole, gates that cannot be undone by the black hole form perturbations. When the size of such perturbation gets large, an infalling observer will see a firewall. Equation \eqref{complexity_perturbation} shows that such perturbation getting large is equivalent to complexity decreasing with time. 

This provides an intuitive explanation for the encounter of a firewall if one breaks the entanglement across the horizon: once the entanglement is disrupted, there will be gates stored in the interior that cannot be undone by the black hole. This argument doesn't explain what happens when one has wrong entanglement across the horizon \cite{Almheiri:2012rt,Almheiri:2013hfa,Marolf:2013dba}.

\section{Operator size and momentum}
\label{size_momentum_in}

In this section we discuss a particular notion of operator growth associated with the increase of momentum of a particle approaching the horizon in the interior.

\subsection{Operator size and exterior momentum}


In \cite{Susskind:2018tei,Brown:2018kvn} it was pointed out that the growth of the perturbing operator $\psi_1$ corresponds to the increase of the dual particle's momentum as it falls toward the black hole from the exterior. More precisely, the operator size in the SYK model is related to exterior momentum in an AdS${}_2$ geometry as:
\begin{align}
\label{AdS2_1}
n_\beta^\text{ex}[\psi_1(t)] \sim P_\text{ex}(t)\tilde\beta \,.
\end{align}
In this equation, $n_\beta^\text{ex}[\psi_1(t)]$ was defined in \eqref{size_exterior}. It represents the average number of fermions in the operator $\psi_1(t)$ in a background of $\rho^{\frac{1}{2}}$ \cite{Qi:2018bje}. The operator $\psi_1(t)$ produces an infalling particle in the dual bulk geometry. On the right hand side of equation \eqref{AdS2_1}, $P_\text{ex}$ is the radial momentum of the particle.\footnote{For a definition of $P$ in general dimensions, see Appendix A of \cite{Brown:2018kvn}.} The local energy scale $\tilde T = \frac{1}{\tilde\beta}$ depends on the radial location of the particle \cite{Brown:2018kvn}.


Earlier work was mostly concerned with particle momentum in the black hole exterior. To be more concrete, in JT gravity the momentum was defined in the following way \cite{Lin:2019qwu}. Say, the particle comes in from the left boundary at time $0$. We look at a time slice anchored at $(t_L, t_R) = (t, -t)$, and ask about its momentum on that slice. These geodesic slices stay mostly in the black hole exterior. The momentum is given by $P \sim N (\partial_{t_L}-\partial_{t_R})L$ where $L$ is the distance between the two boundaries. With this definition, one can show that in JT gravity, the momentum of the particle in the exterior grows as follows (see Appendix \ref{momentum_calculations}):
\begin{align}
\label{momentum_exterior}
	P_\text{ex}(t) \sim N\frac{2\pi}{\beta}\frac{\frac{\delta S}{S-S_0}\sinh\frac{2\pi}{\beta}t}{1+\frac{\delta S}{S-S_0}\qty(\cosh(\frac{2\pi}{\beta}t)-1)}
\end{align}
where $\delta S$ is the increase of the thermal entropy from the perturbation.

Interestingly, $P_\text{ex}(t)$ saturates at a value independent of the initial energy when the particle reaches the horizon, which corresponds to the operator growing to maximal size. In Figure \ref{size_increase}(a), we show the different slices and the direction of size increase. Since these slices don't enter the interior region, it is clear that we will need a different slicing to probe interior properties.

\subsection{Operator size and interior momentum}

If the relation between size and momentum is general \cite{Susskind:1993aa}, what about the momentum of the particle when it's in the interior? If we trace the particle backward from the deep interior toward the horizon, we will again see that its momentum grows exponentially in time.\footnote{A simplified argument in near-Rindler space is given in Appendix \ref{momentum_generalD}.} Do we have a explanation for this momentum growth?

 \begin{figure} 
 \begin{center}                      
      \includegraphics[width=4in]{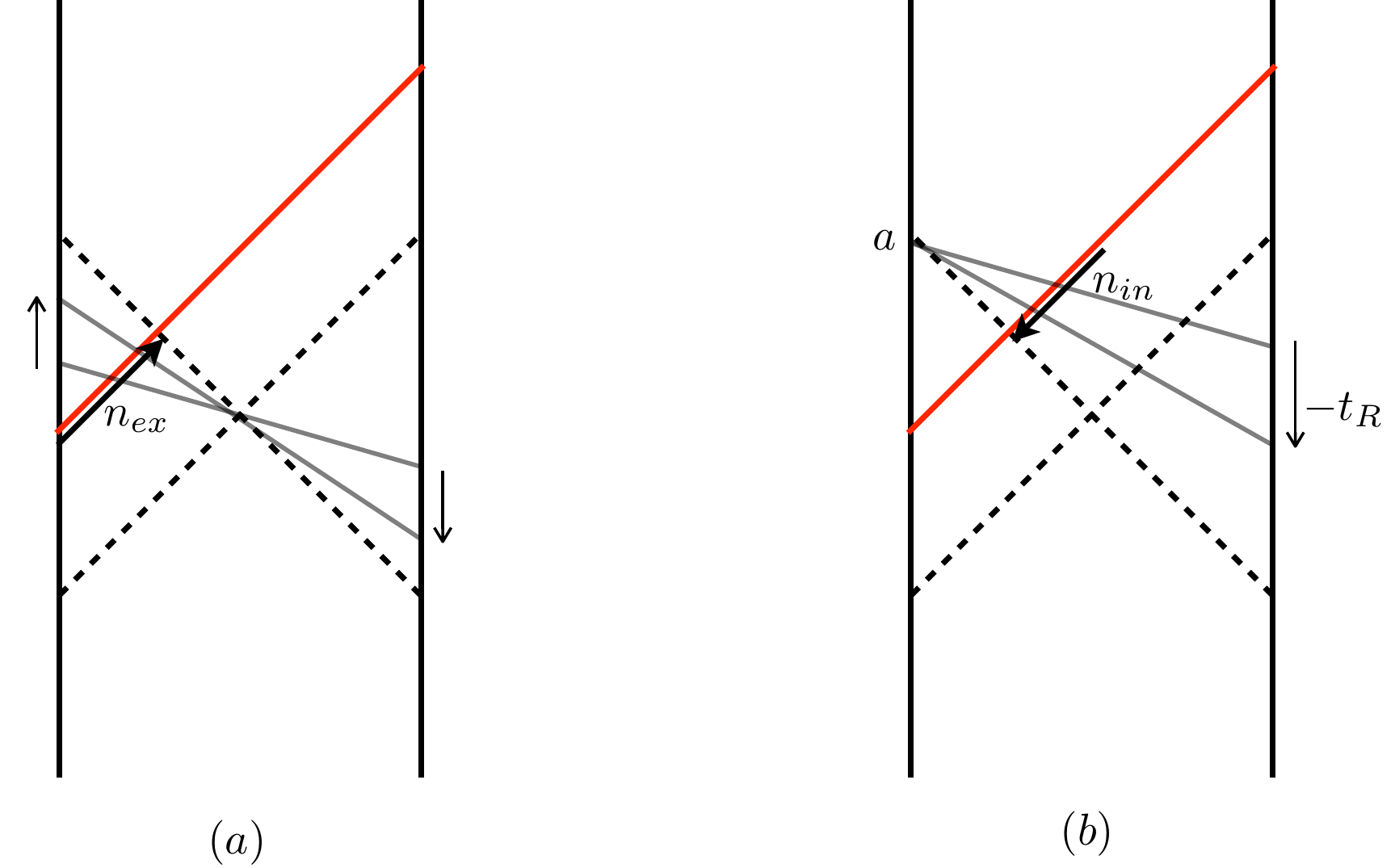}
      \caption{Illustration of time slices used to define (a) exterior momentum, and (b) interior momentum. The arrows show the direction of size increase.}
  \label{size_increase}
  \end{center}
\end{figure}

To make the setup more precise, consider the slicing of AdS${}_2$ shown in Figure \ref{size_increase}(b). We send in a particle from the left boundary at time $0$. To define the interior momentum we look at bulk slices at time $(t_L, t_R) = (a, t_R)$ where $t_*<a\ll \frac{\beta}{2\pi}(S-S_0)$. These slices will intersect the trajectory of the infalling particle in the interior. An evaluation of the momentum in JT gravity (presented in Appendix \ref{momentum_calculations}) yields:
\begin{align}
	\frac{P_\text{in}(a, t_R)}{N\frac{2\pi}{\beta}}  \sim\ & \frac{\frac{\delta S}{S-S_0}\qty(\frac{\sinh(\frac{2\pi}{\beta}\frac{a-t_R}{2})}{\cosh(\frac{2\pi}{\beta}\frac{a+t_R}{2})}+\tanh(\frac{2\pi}{\beta}\frac{a+t_R}{2}))}{1+\frac{\delta S}{S-S_0}\qty(\frac{\cosh(\frac{2\pi}{\beta}\frac{a-t_R}{2})}{\cosh(\frac{2\pi}{\beta}\frac{a+t_R}{2})}+\frac{2\pi}{\beta}a\tanh(\frac{2\pi}{\beta}\frac{a+t_R}{2})-1)}\approx 
	\frac{\frac{\delta S}{S-S_0}(e^{-\frac{2\pi}{\beta}t_R}+1)}{1+\frac{\delta S}{S-S_0}(e^{-\frac{2\pi}{\beta}t_R}+\frac{2\pi}{\beta}a-1)}\label{momentum_interior}
\end{align}
We see that as we fix large $a$ ($t_*<a\ll\frac{\beta}{2\pi}(S-S_0)$) and decrease $t_R$, the momentum grows exponentially in $-t_R$ until saturation at $-t_R = t_*$. It saturates at the same value as the exterior momentum in \eqref{momentum_exterior}.

How is this related to the earlier discussion of the quantum circuit in the interior? If we relate the length of a slice to complexity, we have
\begin{align}
\label{momentum_complexity}
	P \sim \frac{d\mathcal{C}}{dt_L}-\frac{d\mathcal{C}}{dt_R}
\end{align}

At large $a$, we have $\frac{d\mathcal{C}}{dt_L} \sim n_{max}T$ where $T$ is the temperature. We have also seen in earlier sections that $\frac{d\mathcal{C}}{-dt_R} \sim n_{max}T(1-2\mathcal{F})$ on the chosen slices where $\mathcal{F}$ is the percentage of the number of healthy gates as a function of $t_R$. Together with \eqref{momentum_complexity}, we have
\begin{align}
	P_\text{in} &\sim n_{max}T(1+1-2\mathcal{F}) = 2n_{max}T(1-\mathcal{F}) = 2T \, n_\beta^\text{in} \,.
\end{align}
We again see that momentum is related to size ($n_\beta^\text{in} \sim P_\text{in}\beta$), but this time, it is the interior momentum which is related to size defined in \eqref{size_interior}. In Figure \ref{size_increase}(b), the arrow represents the direction of momentum and size increase. The momentum and size again reach their maximal values when the particle approaches the horizon. 

Let's check this correspondence between size and momentum explicitly. If we compare the interior size in \eqref{size_interior} and \eqref{percentage_sick} with the interior momentum in \eqref{momentum_interior}, we see that they have the same early time exponential growth ($-t_R<t_*$) and they both saturate at $-t_R = t_*$. 

\begin{figure} 
 \begin{center}                      
      \includegraphics[width=4in]{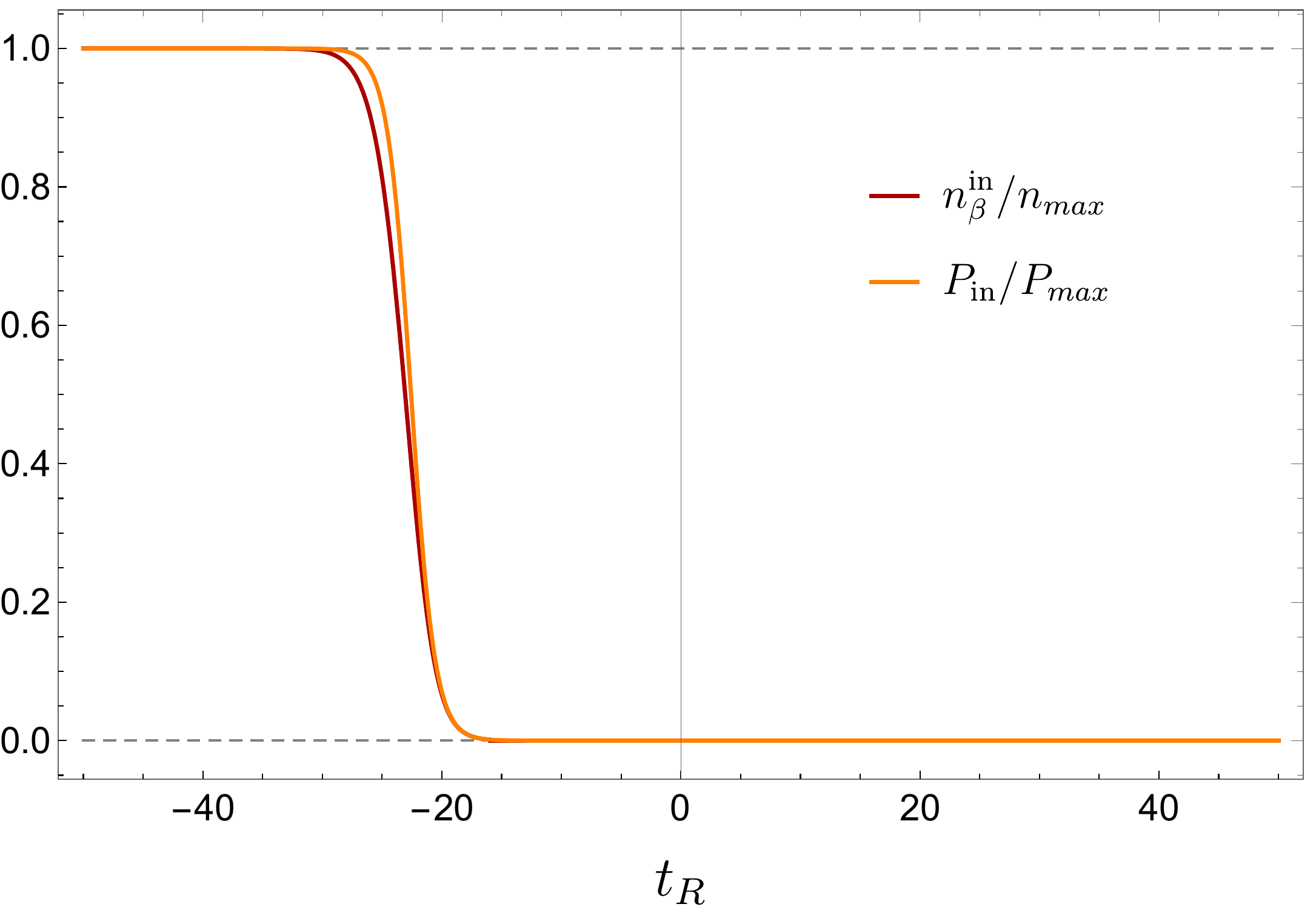}
      \caption{Comparison of the ``interior'' size of $\psi_1(t)$ and the interior momentum as computed in JT gravity. Parameter values are the same as in Figure \ref{FourPtPlot1}.}
  \label{FourPtPlot2}
  \end{center}
\end{figure}


The exact momentum generator in JT gravity was given in \cite{Lin:2019qwu}. Here, we want to emphasize that the interior size defined in \eqref{size_interior} isn't really the exact momentum generator in \cite{Lin:2019qwu}, but has the same qualitative features.\footnote{We discuss the relation between sized defined in \eqref{size_interior} and approximate symmetry generators in Appendix \ref{symmetry_generator}.} We choose to relate interior momentum with \eqref{size_interior} because \eqref{size_interior} has a simple form and a direct relation with size. We expect that it can be generalized to higher dimensions.

\subsection{Interpretation in terms of complementarity}

To summarize, the exterior momentum is related to size defined as \cite{Qi:2018bje}:
\begin{align}
	\frac{n_\beta^\text{ex}[\psi_1(t)]}{n_{max}} = \frac{n_{\infty}[\psi_1(t)\rho^{\frac{1}{2}}]-n_{\infty}[\rho^{\frac{1}{2}}]}{n_{max}-n_{\infty}[\rho^{\frac{1}{2}}]} \,.
\end{align}
It increases as the particle falls from the exterior region towards the horizon, and it saturates at the maximal value at the horizon.
 
The interior momentum is related to size defined in a slightly different way:\footnote{See footnote \ref{alternative_def}.}
\begin{align}
	\frac{n_\beta^\text{in}[\psi_1(t)]}{n_{max}} = \frac{n_{\infty}[e^{iH(a-t)}\psi_1(t)\rho^{\frac{1}{2}}]-n_{\infty}[e^{iH(a-t)}\rho^{\frac{1}{2}}]}{n_{max}-n_{\infty}[e^{iH(a-t)}\rho^{\frac{1}{2}}]} \,.
\end{align}
It also increases as we trace the particle backward from the interior toward the horizon, and it saturates at the same value as the exterior size when reaching the horizon (c.f., Figure \ref{size_increase}).

We want to emphasize that the increase in size on both sides of the horizon does not mean that there is a process during which size increases then decreases. There is only one process involved, that is the growth of the operator $\psi_1(t) = e^{iHt}\psi_1e^{-iHt}$. When we consider its growth on the background of $\rho^{\frac{1}{2}}$, we get the exterior momentum, while if we consider its growth on the background of $e^{iH(a-t)}\rho^{\frac{1}{2}}$, we get the interior momentum. The fact that both exterior and interior momentum are related to slightly distinct notions of operator growth, is a manifestation of complementarity \cite{Susskind:1993if}.

\section{Conclusion and Discussion}
\label{conclusion}

Motivated by the quantum circuit model for black hole evolution, we considered boundary quantities detecting properties of the circuit stored in the black hole interior. In particular, in the case of the perturbed theormofield double state, we proposed that a certain two-sided four-point function counts the fraction of healthy (unaffected) gates in the quantum circuit as a function of time. We connected properties of the quantum circuit with the time dependence of complexity and argued that a lack of healthy gates gives rise to a firewall. We also made a connection between the increase of momentum of a particle in the interior and a suitable notion of operator size. This is complementary to the previously studied increase of exterior momentum.

There are many unanswered questions. We only discussed the part of the particle trajectory close to the outer horizon, which does not detect the region close to the inner horizon.\footnote{For recent work in this direction, see \cite{Lensky:2020fqf}.}

We related the interior perturbation's momentum with operator size defined in certain way in the context of the perturbed thermofield double. It would be interesting to understand how general this relation is, and whether the associated notion of operator size can be put on more formal grounds. In particular, can one generalize this notion of size to pure state black holes? How would a collision in the interior manifest itself? 

Finally, our quantitative statements were obtained in the low-energy sector of the SYK model and through computations in JT gravity. It would be interesting to generalize these ideas to higher dimensions. For example, it would be fascinating to make a precise connection between operator growth and complexity in the context of two-dimensional CFTs.\footnote{This could be achieved, e.g., by extending the framework of \cite{Caputa:2018kdj}.}

\section*{Acknowledgements}
We thank Ahmed Almheiri, Henry Lin, Juan Maldacena, Xiaoliang Qi, Alex Streicher, and Leonard Susskind for helpful discussions. F.H.\ gratefully acknowledges support from the DOE grant DE-SC0009988. Y.Z.\ is supported by the Simons foundation through the It from Qubit Collaboration.

\appendix

\section{Computation of two-sided four-point function}
\label{app:SchwCalc}

In this appendix we review the computation of the quantity \eqref{healthy_1}:
\begin{align}
\label{healthy_1repeat}
	\mathcal{F}(a, t_R) = - \frac{\sum_j \tr(\psi_1 \psi_j(-a)\psi_1\rho^{\frac{1}{2}}\psi_j(t_R)\rho^{\frac{1}{2}} )}{\sum_j \tr(\psi_1\psi_1\rho)\tr(\psi_j(-a)\rho^{\frac{1}{2}}\psi_j(t_R)\rho^{\frac{1}{2}})}
\end{align}
To leading order in large $N$, this quantity is $1$.
In the SYK model in the limit where $N,\beta J \gg 1$ it then receives two types of corrections: $(i)$ a correction due to the exchange of conformal representations, and $(ii)$  another contribution which breaks conformal symmetry, is enhanced by a factor of $\beta J$, and is described by the Schwarzian action. We will only consider the second kind, as this corresponds to the JT gravity analysis. To compute it, we consider generic operators $V$ and $W$ with dimension $\Delta$, which have a dual description in terms of free fields coupled to gravity. We will later fix $\Delta$ such as to match with the SYK considerations. The four-point function of $V$ and $W$ is computed by the integral of bilocal operators over reparametrizations \cite{Maldacena:2016upp}:
\begin{equation}
\label{eq:SchwInt}
    {\cal F}(a,t_R) \propto \int [{\cal D} f]  \; {\cal B}(\hat{u}_1,\hat{u}_2) {\cal B}(\hat{u}_3,\hat{u}_4)\, e^{iS[f]} \,,\qquad 
     {\cal B}(\hat{u}_a,\hat{u}_b) = \left( \frac{ - f'(\hat{u}_a) f'(\hat{u}_b)}{(f(\hat{u}_1)-f(\hat{u}_2))^2} \right)^\Delta
\end{equation}
where $S[f]$ is the Schwarzian action:
\begin{equation}
    S[f] = -C \int dt \, \{ f(t),t \} \,, \qquad C = \frac{S-S_0}{4\pi^2}\, \beta  \,,
\end{equation}
where $S-S_0$ is the near-extremal entropy.
In this section we take $\beta = 2\pi$. We will be interested in the configuration 
\begin{equation}
\label{eq:config}
    \hat{u}_1 = -i(\pi+\delta) \,,\qquad \hat{u}_2 = -i(\pi -\delta) \, \qquad
    \hat{u}_3 = -a - i\pi \,,\qquad \hat{u}_4 = t_R \,.
\end{equation}
The small Euclidean separation $\delta$ serves as a UV regulator. We can either expand resulting expressions in it (and consider the leading divergent terms), or keep it finite (which corresponds to smearing operators slightly).

\paragraph{Single Schwarzian mode exchange:}
We can gain some intuition for the early-time behavior of ${\cal F}(a,t_R)$ by computing the exchange of a single Schwarzian mode $\epsilon$. We can simply work in Euclidean signature and analytically continue in the end. The intermediate result will be the Euclidean out-of-time-order four-point function (see, for instance, eq.\ (4.32) of \cite{Maldacena:2016upp}), which can then be analytically continued to the configuration \eqref{eq:config}:
\begin{equation}
\label{eq:Fapprox}
    {\cal F}(a,t_R) = 1 - \frac{\Delta^2}{C} \frac{1}{\sin \delta} \left[ \frac{2 \sinh \frac{a}{2} \, \sinh \frac{-t_R}{2}}{\cosh \frac{a+t_R}{2}} + a \, \tanh \frac{a+t_R}{2} \right] + \ldots
    \approx 1 - \frac{\Delta^2}{C} \frac{(a + e^{-t_R})}{\sin \delta}+ \ldots
\end{equation}
where we dropped terms of order ${\cal O} \left( C^{-2} \right)$ and gave an approximate form for $a\gg -t_R$. For early times ($\log C \gg -t_R$) we observe the exponential increase in $-t_R$, while for positive $t_R$ (i.e., $C \gg a$ and $t_R>0$) the correlator is approximately $1$.

\paragraph{Higher order resummation:}
In \cite{Maldacena:2016upp} (see also \cite{Shenker:2014cwa,Chen:2016cms,Lam:2018pvp}) it was shown how to improve the above calculation by resumming all relevant contributions for times $a \sim \log C$. The idea is essentially to replace the infinitesimal reparametrization $\epsilon$ by a finite $SL(2,\mathbb{R})$ transformation, which only acts on the sheets of the out-of-time-order contour that feel the propagation of shockwaves created from both pairs of operators simultaneously. The nonlinear Schwarzian action then becomes a simple product of shockwave strengths, and the bilocal operators simplify accordingly. The integral \eqref{eq:SchwInt} can then be performed explicitly. The result reads as follows in our conventions and setup:
\begin{equation}
    {\cal F}(a,t_R) = z^{-2\Delta} \, U(2\Delta,1,\frac{1}{z}) \,,\qquad z = \frac{1}{4C} \frac{1}{\sin \delta} \, \frac{e^{a}}{1+ e^{a+t_R}}\,.
\end{equation}
For $a\gg \log C \gg t_R $ this has the same properties as \eqref{eq:Fapprox}.

%

\section{Momentum calculations in JT gravity}
\label{momentum_calculations}

In this section, we compute the momentum of the infalling particle in JT gravity following \cite{Lin:2019qwu}. The momentum generator on a slice $(t_L, t_R)$ is given by
\begin{align}
	P \sim N (\partial_{t_L}-\partial_{t_R})L
\end{align}
where $L$ is the geodesic distance between the two boundaries in units of the radius of $AdS_2$.
 Say, we send in a particle at left time $0$. $\delta S$ is the change of thermal entropy. We work in embedding coordinates of $AdS_2$. See the Appendix of \cite{Maldacena:2017axo} for details. After the extra particle comes in $(t_L>0)$, we have
\begin{align}
	X_L(t_l)\cdot X_R(t_r)
	 \approx\ & \frac{\beta^2}{4\pi^2}\qty(1-\frac{\delta S}{S-S_0})\qty(-1-\cosh(\frac{2\pi}{\beta}(t_R+t_L)))\nonumber\\
	 &\ \ \ \ \ \ +\frac{\beta^2}{4\pi^2}\frac{\delta S}{S-S_0}\frac{2\pi}{\beta}t_L\qty(-\sinh(\frac{2\pi}{\beta}(t_R+t_L)))\nonumber\\
	 &\ \ \ \ \ \ +\frac{\beta^2}{4\pi^2}\frac{\delta S}{S-S_0}\qty(-\cosh(\frac{2\pi}{\beta}t_L)-\cosh(\frac{2\pi}{\beta}t_R))
\end{align}
The geodesic distance between the two sides is
\begin{align}
	L	=\ &\log(-2\frac{X_L(t_L)\cdot X_R(t_R)}{\epsilon^2})\nonumber\\
	=\ &\log(\frac{\beta^2}{\pi\epsilon^2})+\log(\cosh^2\qty(\frac{2\pi}{\beta}(\frac{t_L+t_R}{2})))\nonumber\\
	&+\log\qty[1+\frac{\delta S}{S-S_0}\qty(\frac{\cosh(\frac{2\pi}{\beta}\frac{t_L-t_R}{2})}{\cosh(\frac{2\pi}{\beta}\frac{t_L+t_R}{2})}+\frac{2\pi}{\beta}t_L\tanh(\frac{2\pi}{\beta}\frac{t_L+t_R}{2})-1)]
\end{align}
which gives the following expression for momentum:
\begin{align}
	P \sim\ & N(\partial_{t_L}-\partial_{t_l})L = N\frac{2\pi}{\beta}\frac{\frac{\delta S}{S-S_0}\qty(\frac{\sinh(\frac{2\pi}{\beta}\frac{t_L-t_R}{2})}{\cosh(\frac{2\pi}{\beta}\frac{t_L+t_R}{2})}+\tanh(\frac{2\pi}{\beta}\frac{t_L+t_R}{2}))}{1+\frac{\delta S}{S-S_0}\qty(\frac{\cosh(\frac{2\pi}{\beta}\frac{t_L-t_R}{2})}{\cosh(\frac{2\pi}{\beta}\frac{t_L+t_R}{2})}+\frac{2\pi}{\beta}t_L\tanh(\frac{2\pi}{\beta}\frac{t_L+t_R}{2})-1)}
\end{align}

\section{Interior momentum}
\label{momentum_generalD}
\begin{figure}
 \begin{center}                      
      \includegraphics[width=1.8in]{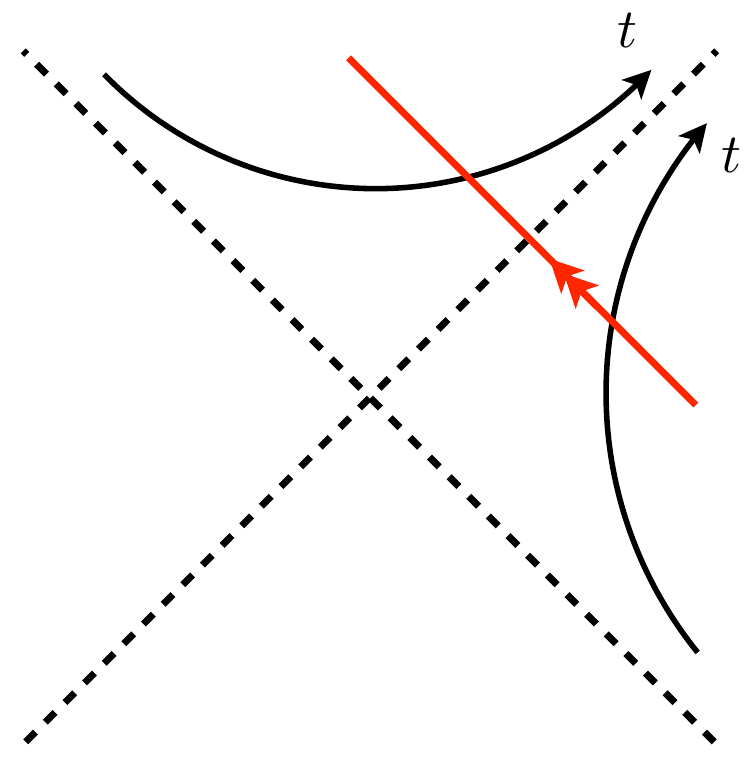}
      \caption{Infalling particle in Rinder space.}
  \label{momentum_general}
  \end{center}
\end{figure}

In this section we show that the momentum of an infalling object grows exponentially in time as we trace it backward from the interior toward the horizon. We write the interior metric as (c.f., Figure \ref{momentum_general})
\begin{align}
    ds^2 =\ & \frac{dr^2}{f(r)}-f(r)dt^2+r^2d\Omega_{D-2}^2\nonumber \\
    =\ & -d\rho^2+h(\rho)dt^2+g(\rho)d\Omega_{D-2}^2 \,,
\end{align}
where $\rho$ is the radial proper time: $d\rho = \frac{dr}{\sqrt{-f(r)}}$. $h(\rho) = -f(r)$ and $g(\rho) = r^2$. 

Let $\tau$ be the proper time of the infalling particle. The momentum conjugate to $\rho$ is given by $\frac{d\rho}{d\tau}$. For massless particle, it is given by
\begin{align}
    P_\text{in} = \frac{d\rho}{d\tau} = \frac{E}{\sqrt{h(\rho)}} = \frac{E}{\sqrt{-f(r)}}
\end{align}
where $E$ is the conserved momentum conjugate to $t$. 

The particle trajectory satisfies $dt = -\frac{dr}{f(r)}$. If we approximate the near-horizon metric by Rindler geometry: $f(r) = \frac{4\pi}{\beta}(r-r_h)$, we have
\begin{align}
    &|r_h-r| \propto e^{-\frac{4\pi}{\beta}t} \,, \qquad P_\text{in}\propto E e^{\frac{2\pi}{\beta}t} \,.
\end{align}

\section{Approximate symmetry generators}
\label{symmetry_generator}

We want to show that the size defined in \eqref{size_interior} is a good approximation to momentum. We first replace $e^{iH(a-t)}$ by $e^{iHa}$, see footnote \ref{alternative_def} and get
\begin{align}
    1 - \frac{n_\beta^\text{in}[\psi_1(t)]}{n_{max}} 
    =\ &\frac{\sum_{j = 1}^N\bra{\text{TFD}}\psi_1^L(-t)e^{iH_La}\psi_j^L\psi_j^Re^{-iH_La}\psi_1^L(-t)\ket{\text{TFD}}}{\sum_{j = 1}^N \bra{\text{TFD}}e^{iH_La}\psi_j^L\psi_j^Re^{-iH_La}\ket{\text{TFD}}}
\end{align}
We consider the coupled Hamiltonian
\begin{align}
	&\hat{E} = H_L+H_R+i\mu\sum_j\psi_L^j\psi_R^j-\expval{H_L+H_R+i\mu\sum_j\psi_L^j\psi_R^j}_{e^{-iH_La}\ket{\text{TFD}}}
\end{align}
From \cite{Maldacena:2018lmt}, the ground state of this Hamiltonian is $\ket{\text{TFD}}$ for appropriate $\beta(\mu)$. We also have the boost generator $\hat{B} = H_L-H_R$.
Their sum generates the boost around the right boundary point $t_R = 0$:
\begin{align}
	\hat{E}+\hat{B} = 2H_L+i\mu\sum_j\psi_L^j\psi_R^j-\expval{H_L+H_R+i\mu\sum_j\psi_L^j\psi_R^j}_{e^{-iH_La}\ket{\text{TFD}}}
\end{align}
The expectation value of this quantity is related to the ``interior'' size:
\begin{align}
	&\bra{\text{TFD}}\psi_1^L(-t)e^{iH_La}(\hat{E}+\hat{B})e^{-iH_La}\psi_1^L(-t)\ket{\text{TFD}}\nonumber\\
	 &\quad = \bra{\text{TFD}}\psi_1^L(-t)e^{iH_La}\qty(i\mu\sum_j\psi_L^j\psi_R^j)e^{-iH_La}\psi_1^L(-t)\ket{\text{TFD}}-\expval{i\mu\sum_j\psi_L^j\psi_R^j}_{e^{-iH_La}\ket{\text{TFD}}}\nonumber\\
	 &\qquad +\bra{\text{TFD}}\psi_1^L(-t)(2H_L)\psi_1^L(-t)\ket{\text{TFD}}-\bra{\text{TFD}}(2H_L)\ket{\text{TFD}}
\end{align}
The first line is proportional to $n_{max}-n_{\beta}^{in}$, while the second line doesn't grow with $t$. 
Note that $\hat{E}$ and $\hat{B}$ here are only approximate generators. They differ from the exact symmetry generators defined in \cite{Lin:2019qwu}. As $\hat{E}^2-\hat{B}^2-\hat{P}^2$ stays constant and $\hat{B}$ doesn't grow, the time dependence of $\hat{E}+\hat{B}$ and $\hat{P}$ are roughly the same.

\bibliographystyle{apsrev4-1long}
\bibliography{reference}

\end{document}